\documentclass[10pt,a4paper]{article}

\usepackage{graphicx}
\usepackage{amsmath,amsfonts}%
\usepackage{tabularx}
\usepackage{enumitem}
\usepackage{xspace}
\usepackage{authblk}
\usepackage{multirow}

\def\A{\mathcal{A}}

\def\H{\mathcal{H}}

\def\ie{{\it i.e..,\xspace}}

\title{Consciousness as a global property of brain dynamic activity}

\author[1,*]{D.M. Mateos }
\author[2]{R.Wennberg}
\author[3]{R. Guevara}
\author[1,4]{J.L. Perez Velazquez}
\affil[1]{\normalsize Neuroscience and Mental Health Programme, Division of Neurology, Hospital for Sick Children.\\
Institute of Medical Science and Department of Paediatrics, University of Toronto, Toronto, Canada. }
\affil[2]{\normalsize Krembil Neuroscience Centre, Toronto Western Hospital, University of Toronto, Toronto, Canada.}
\affil[3]{\normalsize Laboratoire Psychologie de la Perception, CNRS and Universite Paris Descartes, Sorbonne Paris Cite, Paris, France.}
\affil[4]{\normalsize Ronin Institute, Montclair, USA}

\affil[*]{Corresponding author: Diego M. Mateos, mateosdiego@gmail.com .}
\begin{document}

\maketitle

\begin{abstract}
  We seek general principles of the structure of the cellular collective activity associated with conscious awareness. Can we obtain evidence for features of the optimal brain organization that allows for adequate processing of stimuli and that may guide the emergence of cognition and consciousness? Analysing brain recordings in conscious and unconscious states, we followed initially the classic approach in physics when it comes to understanding collective behaviours of systems composed of a myriad of units: the assessment of the number of possible configurations (microstates) that the system can adopt, for which we use a global entropic measure associated with the number of connected brain regions. Having found maximal entropy in conscious states, we then inspected the microscopic nature of the configurations of connections using an adequate complexity measure, and found higher complexity in states characterised not only by conscious awareness but also by subconscious cognitive processing, such as sleep stages. Our observations indicate that conscious awareness is associated with maximal global (macroscopic) entropy and with the short time scale (microscopic) complexity of the configurations of connected brain networks in pathological unconscious states (seizures and coma), but the microscopic view captures the high complexity in physiological unconscious states (sleep) where there is information processing. As such, our results support the global nature of conscious awareness, as advocated by several theories of cognition. We thus hope that our studies represent preliminary steps to reveal aspects of the structure of cognition that leads to conscious awareness.
\end{abstract}


\section{Introduction}
\label{Introduction:sec}

While the coordinated patterns of organized activity in the brain that are associated with cognitive states are definitely intricate, the underlying global principles may be simple, despite those mechanistically complicated processes. The current overabundance of data should not preclude attempts at a comprehension of fundamental principles. Are there minimal, basic conditions for the nervous system to satisfy to optimally process information? Although the structure of anatomical connectivity sets limits to information transfer among cell ensembles, there could be a principle governing the activity of these networks to optimize the integration and segregation of information associated with conscious awareness.  There have been proposals whose underlying theme is that ongoing transformation of information in the brain is reflected in the variability and fluctuations of the widespread functional connections among brain cell ensembles that manifest in aspects of consciousness. At the same time, it has been proposed that consciousness requires a certain high complexity in the organization of coordinated activity of brain cell ensembles. There are many notions of complexity, and structural as well as effective measures of complexity have been advanced \cite{adami2002complexity}, with the view that functional organization is determined and superimposed on the structural organization \cite{chauvet1993non}. The question with regards to the nervous system is to find the most adequate notion that captures its “complexity”. Starting from the most basic neurophysiological aspects, it is known that the \textit{coordinated collective activity} of the constituent cells (neurons and glial cells) is the essential aspect that ensures a proper performance to aid the organism in responding to a changing environment. The key elements are not individual cells but networks, or ensembles of interacting cells. The activity of nervous system cells is to a large extent coherent, showing high degrees of temporal correlation/coordination of activity such that cell assemblies self-organise via transient synchronization \cite{singer2006phenomenal,varela2001brainweb}. A prominent question is how to describe the organizing principles of this collective activity which allow features associated with consciousness to emerge. Therefore, it is probably in the scrutiny of the correlations of activity where answers to those questions can be found. Recent studies have in fact sought simplicity in the principles governing the number of interactions among variables describing networks \cite{stephens2011searching}. In our study we focus on the collective level of description and assume that coordinated patterns of brain activity evolve due to interactions of mesoscopic areas that can be recorded electrophysiologically.

A previous study showed that the specific values of a synchrony index (see Methods) obtained in fully alert states represent the largest number of configurations of pairwise signal combinations and thus have higher entropy \cite{erra2016statistical}. Those results were obtained from a global perspective, as the calculation of the entropy in each cognitive state represented the total number of configurations of connected signals over a macroscopic time scale lasting several minutes. Based on the same notion of pairwise connected signals, we have performed here an estimation of the joint Lempel--Ziv complexity (JLZC) that allows for a microscale perspective of the fluctuations in the connectivity between two signals, as described in Methods. In this study we have focused on this method and have only computed the entropy to illustrate the differences (or similarities) in the values of entropies and the JLZC, and in this manner we avoid a duplication of the data presented in \cite{erra2016statistical}. As explained in Methods, we have used clinical neurophysiologic data obtained in individuals in different mental states and recorded with three methods: scalp EEG, intracerebral EEG (iEEG), and magnetoencephalography (MEG). To facilitate the comparison of the results for all cases, the normalized joint Lemplel-Ziv Complexity (nJLZC) is shown. One reason we wished to use distinct recording techniques was to ascertain that the results were not dependent upon the recording methodologies. 

 \section{Methods}
\label{Method:sec}

\subsection{Electrophysiological recordings}

Recordings were analyzed from 27 subjects. Some patients were described in Guevara et al. \cite{guevara2005phase}. Specifically, two patients with different epilepsy syndromes were studied with MEG; one patient with temporal lobe epilepsy was studied with iEEG; 3 patients with frontal or temporal lobe epilepsy were studied with simultaneous iEEG and scalp EEG; and 4 nonepileptic subjects were studied with scalp EEG.

For the study of seizures versus alert states, the 2 subjects with MEG recordings and the temporal lobe epilepsy patient investigated with iEEG were used. Details of these patients' epilepsies, seizure types and recording specifics have been presented in previous studies (MEG patients in \cite{dominguez2005enhanced}; iEEG patient in \cite{velazquez2011experimental}). In addition to our patients, we also used iEEG data in 7 other patients obtained from the European Epilepsy Database \cite{ihle2012epilepsiae}. The database contains well-documented meta data, highly annotated raw data as well as several other features. Acquisition rates varied from 254 to 1024 Hz and these differences were taken into consideration for the data analyses. For more information about the recordings  and the settings see  \cite{ihle2012epilepsiae}. For the study of sleep versus alert states, scalp EEG recordings from 5 subjects were used. The 3 patients with combined scalp EEG-iEEG have been described previously in \cite{wennberg2010intracranial}; the two subjects with only scalp EEG were investigated for syncope, with no evidence of epilepsy found during prolonged monitoring. For the study of coma, scalp EEG recordings from two patients in coma secondary to hypoxic brain injury were used, alpha coma in one case  and burst-suppression pattern in the other.

MEG recordings were obtained using a whole head CTF MEG system (Port Coquitlam, BC, Canada) with sensors covering the entire cerebral cortex, whereas iEEG subdural and depth electrodes were positioned in various locations in the frontal and temporal lobes depending on the clinical scenario, including  the amygdala and hippocampal structures of both temporal lobes. EEG, iEEG and EEG-iEEG recordings were obtained using an XLTEK EEG system (Oakville, ON, Canada). Acquisition rates varied from 200 to 625 Hz and these differences were taken into consideration for the data analyses. The duration of the recordings varied as well: for the seizure study,  MEG sample epochs were each of 2 minutes duration, with total recording times of 30-40 minutes per patient; the iEEG patient sample epoch selected for analysis from a continuous 24-hour recording was of 55 minutes duration. The sleep study and coma data segments were each 2-4 minutes in duration, selected from continuous 24-hour recordings or 30 minute clinical recordings (coma).

The EEG control group (10 healthy subjects) were taken from the Physionet \textit{EEG Motor Movement/Imagery Dataset }. The system is described in \cite{schalk2004bci2000, goldberger2000components}

\subsection{Joint Lempel--Ziv complexity of pairwise connections.}


Based on the idea of Kolmogorov   complexity,  Lempel and Ziv developed their algorithmic complexity using the idea of a program based on the recursive copy and paste operation \cite{LemZiv76}.  Their  definition  lies  on  the  two
fundamental notions of reproduction and production. If we consider a  finite size sequence $S_n = s_1 \ldots  s_n$ of symbols of an alphabet $\A$ of  finite size $\alpha = \left| \A  \right|$ it defines:
\begin{itemize}
\item  {\bf Reproduction:}  A  process  of reproduction  from  a sequence  $S_n$
  consists in its extension $R_{n+m} = S_n Q_m$ where $Q_m$  is a sub-sequence
  of length  $m$ of  the sequence  $S_n Q_m \epsilon$,  where $\epsilon$  is the
  operation of suppression  of the last symbol. In other  words, there exists an
  index $p \le  n$ (called pointer) in  the sequence $S_n$ so that  $q_1 = s_p$,
  $q_2 = s_{p+1}$ if  $p < n$ and $q_2 = q_1$  otherwise (the symbol just copied
  in this case), etc. As an example,  $S_3 = 1\, 0 \, 1$ reproduces the sequence
  $R_6 = 1 \, 0 \, 1 \, 0 \, 1  \, 0$ since $Q_3 = 0 \, 1 \, 0$ is a subsequence
  of $S_3 Q_3 \epsilon =  1 \, 0 \, 1 \, 0 \, 1$, \ie $q_1  = s_2 = r_2$, $q_2 =
  s_3 = r_3$ and $q_3 = q_1  = r_4$ previously copied.  In other words, $SQ$ can
  be reproduced from $S$ by recursive copy and paste operations. In a sense, all
  the ``information'' of the extended sequence is in $S_n$.
\item {\bf Production:} A production operation, denoted $S_n \Rightarrow R_{n+m}
  = S_n  Q_m$ consists  in reproducing the  subsequence $R_{n+m}  \epsilon$ from
  $S_n$. The last symbol can  also follow the recursive copy-paste operation, so
  that the  production is a reproduction, but  can be ``new''. Note  thus that a
  reproduction is also  a production, but the converse is  false.  As an example
  $S_3 = 1\, 0 \, 1$ produces $S_3 Q_3 = 1 \, 0 \, 1 \, 0 \, 1 \, 1$ but it does
  not reproduce this sequence: $q_1 = s_2 = r_2$, $q_2 = s_3 = r_3$ but the last
  symbol does  not follow the recursion,  $q_3 \ne r_4$.  The difference between
  reproduction and  production is  that in production  the last letter  can come
  from a supplementary copy-paste but can also be ``new''.
\end{itemize}

The sequence generated is called a History \textit{History} $\H$ and there exist many different histories $\H_i$ for the same sequence. 
Then, for a given $\H_i$  of the sequence, let us define by $C_{\H_i}(S_n)$
the number of productions of  this history.  Clearly $\min(2,n) \le C_{\H_i}(S_n)
\le n$.  In the spirit of  the Kolmogorov complexity, Lempel and Ziv defined the
complexity of the sequence as  the minimal number of production processes needed
to generate it,
\begin{equation}
C(S_n) = \min_{\H_i \in \{\mbox{\tiny histories of $S_n$}\}} C_{\H_i}(S_n)
\end{equation}
It  can  intuitively  be  understood  that  in  the  optimal  history,  all  the  productions are not reproductions, otherwise  it would be possible to reduce the number of steps: the optimal  history is indeed exhaustive \cite{LemZiv76}. This fact  allowed  the  developments  of  simple algorithms  of  evaluation  of  the  Lempel--Ziv complexity of a sequence \cite{KasSch87}. The total number of subsequences present in $S_n$ has an upper
bound \cite{hu2006analysis}, denoted as $L(S_n)$,
\begin{equation}
L(S_n)=C(S_n)[log_\alpha(C(S_n))+1].
\end{equation}
For  large $n$  the normalized Lempel-Ziv complexity is defined as,
\begin{equation}
\label{LZC_normalization}
c(S_n)=\frac{C(S_n)[log_\alpha(C(S_n))]}{n}.
\end{equation}


\subsection{Joint Lempel--Ziv Complexity}

Originally the Lempel Ziv--Complexity (LZC) was used for temporal analysis. Zozor et al.\cite{ZozRav05} proposed, in 2005,  an extension of the method to analyse multidimensional signals, which became therefore useful for spatio-temporal analysis. 
The main idea is  to use the LZC for vectorial data and this can be done naturally extending the alphabet. Consider $k$ 
sequences $\boldsymbol{X}_i=x_{i,1};...;x_{i,n}$ for  $i=1,...,k$ , where the letters belong to the alphabet $\mathcal{A}=\{ 
1,..., \alpha  \}$ \footnote{ In the original paper the authors generalize the method 
using different alphabets for each signal. In our case, the alphabet is the 
same.}. Now, for the sequence $\boldsymbol{V}=v_1,...,v_n$ where the component $v_j$ 
has the $x_{i,j}$ as $\alpha$-ary descomposition, i.e., $v_j=\sum _ {i=1}^{k} 
x_{i,j}~\alpha^i$, the  joint Lempel-Ziv Complexity (JLZC) is defined as 
\begin{equation}
 C(\boldsymbol{X}_1,...,\boldsymbol{X}_{k})=C(\boldsymbol{V})
 \end{equation} 
Using this approach, the algorithm proposed in \cite{KasSch87} to evaluate the LZC can still be
used, comparing scalars.
Moreover, one of the most interesting properties of the JLZC is the invariance  by any permutation $\sigma$ of $\{ 1,...,k\} $, i.e., $C(X_1,...,X_k)=C(X_{\sigma (1)},...,X_{\sigma (k)})$ \cite{ZozRav05}. 

\begin{figure}[htbp]
\includegraphics[scale=0.85]{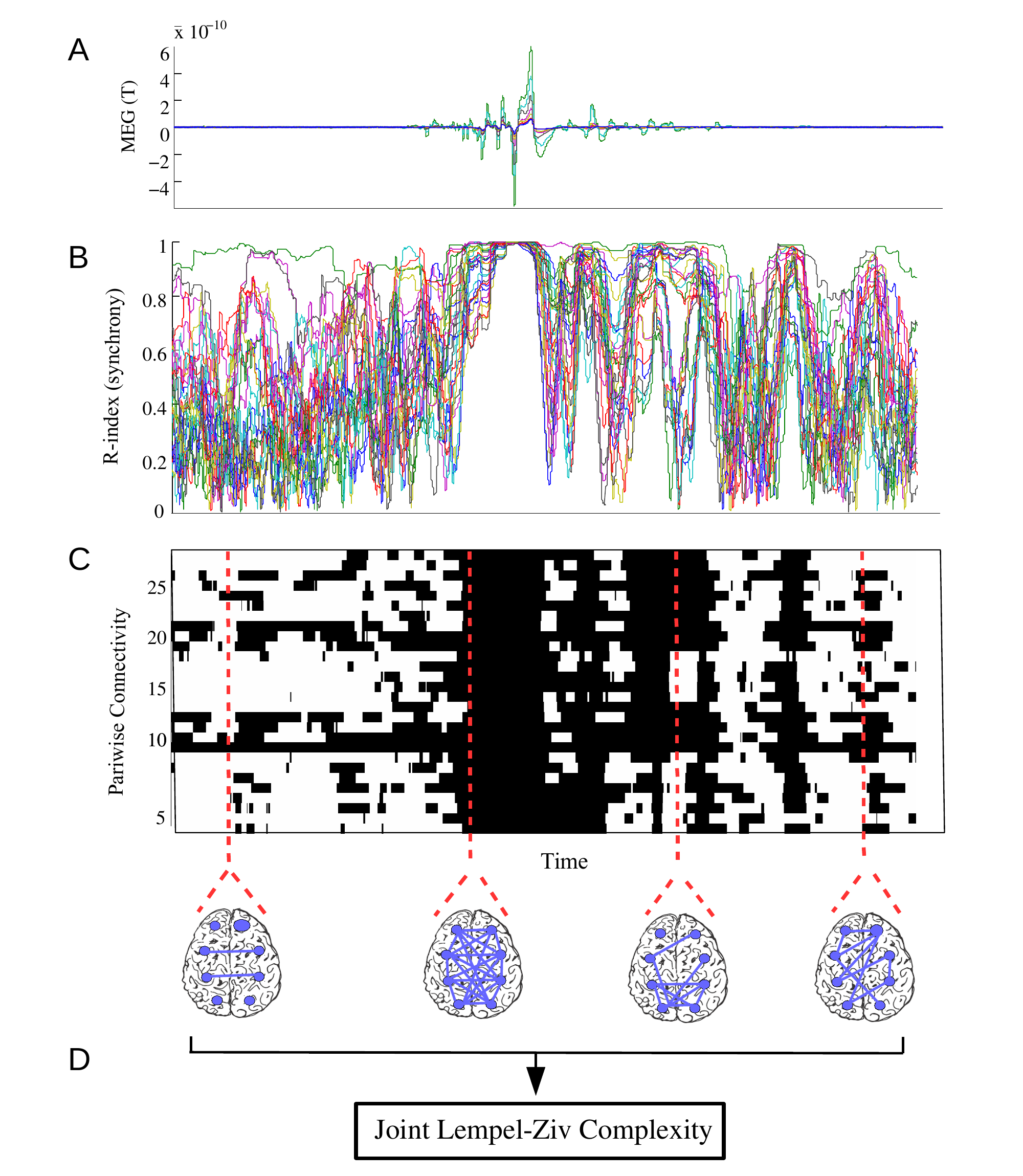} 
\caption{ Joint Lempel Ziv complexity applied over the synchrony index (R). A) Magnetoencephalographic signal recorded in 8 channels. B). The synchony index calculated over all possible pairs ($N_p = 28$). C) Binary connection matrix, black = connected, white=not connected. On the bottom, the schematic brains represent in a pictorial manner what the rows and columns of the binary matrix composed of 1s  and 0s mean (detailed in Methods), D) The fluctuating connections among brain areas that will be assigned a value of JLZC depending on the variability of the connectivity patterns.} 
\label{MethodFig}
\end{figure}

We apply the LZC to our measures of phase synchronization in the following manner (see the schematic in Figure\ref{MethodFig}). Initially a phase synchrony index (R) was calculated from all possible pairwise signal combinations, for which we use the standard procedure of estimating phase differences between two signals from the instantaneous phases extracted using the analytic signal concept via the Hilbert transform. The methods have been extensively described in several publications, so we refer the reader to a few representative \cite{mormann2000mean,dominguez2005enhanced}. In brief, to compute the synchrony index, several central frequencies, as specified in the text, were chosen with a bandpass filter of 2 Hz on either side; hence, for one value of the central frequency $f$ , the bandpass is $f \pm  2$ Hz. The central frequencies were chosen according to the relevant behavioral states and some analytical limitations. To see whether similar results were obtained with different frequencies, we chose several provided there was power at those values (note of caution: if there is very little power at some frequency our methods cannot be applied since they are based on the extraction of phases of oscillations). The phase synchrony index $R$  was calculated using a $t_w=1s$ running window and was obtained from the phase differences using the mean phase coherence statistic which is a measure of phase locking and is defined as $R = | \langle  e^{i\phi} \rangle _{t_w} |$ where $\phi$ is the phase difference between two signals. This analytical procedure has been described in great detail elsewhere \cite{mormann2000mean,dominguez2005enhanced,guevara2005phase}. We note that the only pre-processing of the data was done in the case of the scalp EEG recordings, where a Laplacian derivation was used to decrease volume conduction and the common reference contamination \cite{guevara2005phase,erra2016statistical}.

For each pair of signals we obtain $\{ R(t) \}$ sequences. We have to binarize these continuous sequences 
using a threshold $T_h$, between $0$ (non ``connected'') when $R(t)< T_h$ and $1$ (connected) otherwise. The 
thresholds were obtained from the mean $R$ given by surrogates (we used 10 surrogates per each original signal). Here, as already mentioned in the main text,  we remark again that we are using the term  ``connectivity'' although in reality our synchrony analysis reveals only correlations between phases of oscillations; it is, however, reasonable to infer that these correlations of activity may underlie some degree of connectivity. After the discretization we obtain $N_p$   binary  sequences,  where each row vector $\boldsymbol{B}_i=b_{i,1},...,b_{i,t}$  corresponds to the connectivity between two recording channels.

We then apply the aforementioned JLZC over this matrix. In this case the alphabet length is $\alpha=2$ and the component of the vector  $\boldsymbol{V}$ are  $v_j=\sum _ {i=1}^{N_p} b_{i,j}~2^i$, where $b_{i,j}$ are the components of the binary connection matrix. Finally, to normalize the measure we use the equation \ref{LZC_normalization}. Now, each column of the binary connection matrix represents the brain state at time $t_i$, every state is assigned a letter of the alphabet $\A$; the Lempel-Ziv complexity measures the complexity over the string of these letters (states). If the system remains in the same state the letters of the sequence are the same and thus  JLZC is 0. The same happens if the sequence has fluctuations that repeat themselves, the states/letters that will appear will be concatenated in the same fashion giving low JLZC. But if the state changes all the time without a pattern, the number of the configuration states/letters will be higher giving a higher complexity. Thus JLZC captures information from the system than the entropy (see section \ref{ConecEntropy}) cannot. As an example  if we have two different  binary strings $X_1=0 1 0 1 0 1 0 1$ and $ X_2=1 1 0 1 0 0 0 1$,  the entropy is $S(X_1)=S(X_2)=0.5$ but the complexity  is $C(X_1)=3 $ and $C(X_2)=4$. Another advantage  of JLZC is that it does not require to compute the probability distribution function as the calculation of entropy needs.

For our estimations of JLZC and entropy a complication emerges if the number of signals is large. The total number of possible pairs of channels for a  specific montage is given by the binomial coefficient $N_p = N_c! / 2! (N_{c} - 2)!$ where $N_c$ is the number of channels taken for the analysis. Then we have the $2$-ary decomposition alphabet $N_{\alpha}=2^{N_p}$. If we take $N_c=20$ channels, we have $N_p=190$ pairs of channels and $N_{\alpha}= 1.57x10^{57}$. For $N_c>20$ we have $N_{\alpha}\rightarrow \infty$, and  which is impossible to manage. In this work, to compute the JLZC,  we took $N=20$ for all the montages that have more that 20 channels. To compare the values for different montages, we normalize the Lempel--Ziv complexity applying equation \ref{LZC_normalization}. The invariant  permutation property of the JLZC allows to put each recording in any position of the binary matrix $B$ without changing the JLZC value.   Finally, it is important to point out that the JLZC depends on  the recording sampling frequency,  because the number of possible configurations appearing in a period of time depends on the number of data points. Thus, only signals recorded with similar sampling frequencies should be compared. It is for this reason that in our study we cannot put together results obtained in different patients, and these are shown according to their own specific sampling rates.

\subsection{Entropy of the number of connections}
\label{ConecEntropy}

The calculation of entropy was described in \cite{erra2016statistical}, here we briefly summarise the method. The  number of pairwise connected signals is obtained from the aforementioned threshold of the synchrony index. This  gives us a Boolean connectivity matrix, with 0 entry if the corresponding synchrony index is lower than a  threshold, and 1 if higher, and two channels are “connected” if the corresponding entry in the matrix is 1. The  total number of possible pairs of channels given a specific channel montage is given by the binomial coefficient   
$ N = N_{c}! / 2! (N_{c} - 2)! $  where Nc is the total number of channels in the recording montage. For each subject and 
condition we calculate $p$, the number of connected pairs of signals in the different behavioural states, using the 
aforementioned threshold of the synchrony index, and estimate C, the number of possible combinations of those p 
pairs, using the binomial coefficient again:  $C = N!/ p!~(N-p)!$  In sum, all these calculations represent the 
relatively simple combinatorial problem we are trying to solve: given a maximum total of $N$ pairs of connected 
signals, in how many ways can our experimental observation of p connected pairs (that is, the number of 1’s in the 
matrix) be arranged. To assess entropy we assume that the different pairwise configurations are equiprobable, thus 
the entropy is reduced to the logarithm of the number of states, $S = ln( C)$. However, the estimation of $C$ is 
not feasible due to the large number of sensors; for example, for 144 sensors (our typical MEG montage), the total 
possible number of pairwise connections is $[144~2]=10296$, then if we were to find in the experiment that, say, 
2000 pairs are connected, the computation of $[10296~2000]$ has too large numbers for numerical manipulations. To 
overcome this difficulty, we used the well-known Stirling approximation for large n: $ ln(n!) = n~ln(n) - n$ . The 
Stirling approximation is frequently used in statistical mechanics to simplify entropy-related computations. Using 
this approximation, and after some basic algebra, the equation for entropy reads, $S = N~ln (N / N-p) - p ~ln (p/ N-
p)$ where $N$ and $p$, defined above, are the total number of possible pairs of channels and the number of connected 
pairs of signals in each experiment, respectively. Because this equation is derived from the Shannon entropy, it 
indicates the information content of the system as well \cite{rieke1999spikes}. In the figures, the continuous 
curves shown are obtained from the Stirling approximation equation above, representing the possible entropy values 
of all possible numbers of pairwise combinations, yielding an inverted U. On that curve, we plot the data points 
associated with each specific condition. 

\section{Results}
\begin{figure}[htbp]
\includegraphics[scale=0.8]{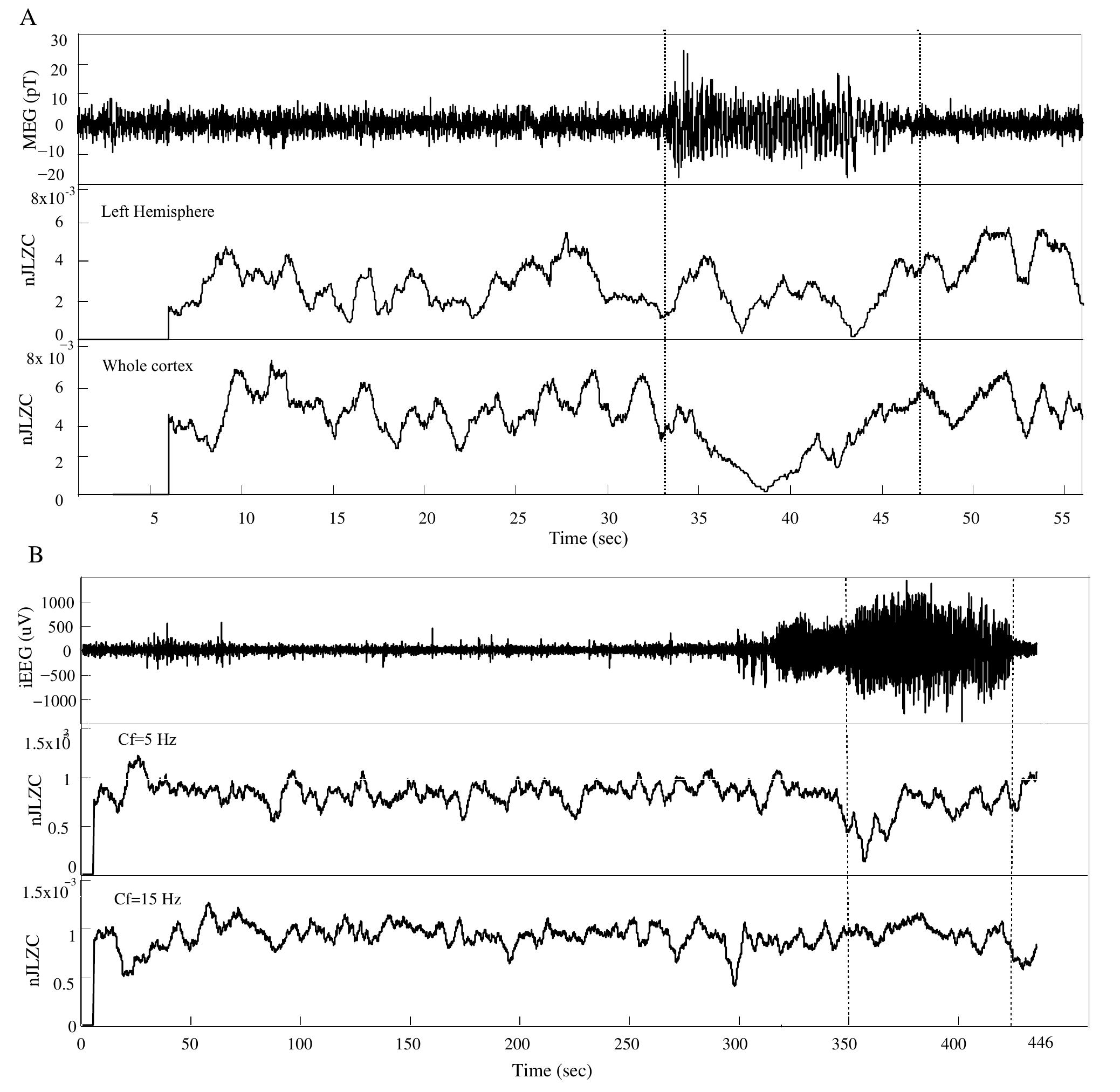} 
\caption{Normalized joint Lempel-Ziv complexity (nJLZC) of the transition between interictal-ictal activity in two epileptic patients. Seizures are marked by the dotted lines. A) Upper graph shows one MEG signal taken in a patient with primary generalized absence epilepsy and seizures associated with loss of consciousness. Middle and lower graphs depict, respectively, the  nJLZC (using a 1-second running window) estimated from several sensors in the right hemisphere and in both hemispheres for a central frequency $Cf=5 Hz$. Note the clearer decrease in complexity when signals from both hemispheres are considered, indicating that seizures in this type of epilepsy (generalized spike-and-wave) are best described as a global state. B) Upper graph shows one iEEG signal taken in a patient with frontal lobe epilepsy. Middle and lower graphs represent the  nJLZC calculated using all intracerebral sensors at the central frequency of $5 Hz$ and $15 Hz$ respectively. It is important to note that in this case the seizure, which started at ${\sim} 320$ s, did not generalize until ${\sim}350$ s according to the clinical report (first dotted line), and it was at this time when the patient lost consciousness. Second dotted line indicates end of the seizure. The drop in the complexity is clear at $5 Hz$ after the seizure generalized. }
\label{fig:Figure1}
\end{figure}

\begin{figure}
\includegraphics[scale=0.85]{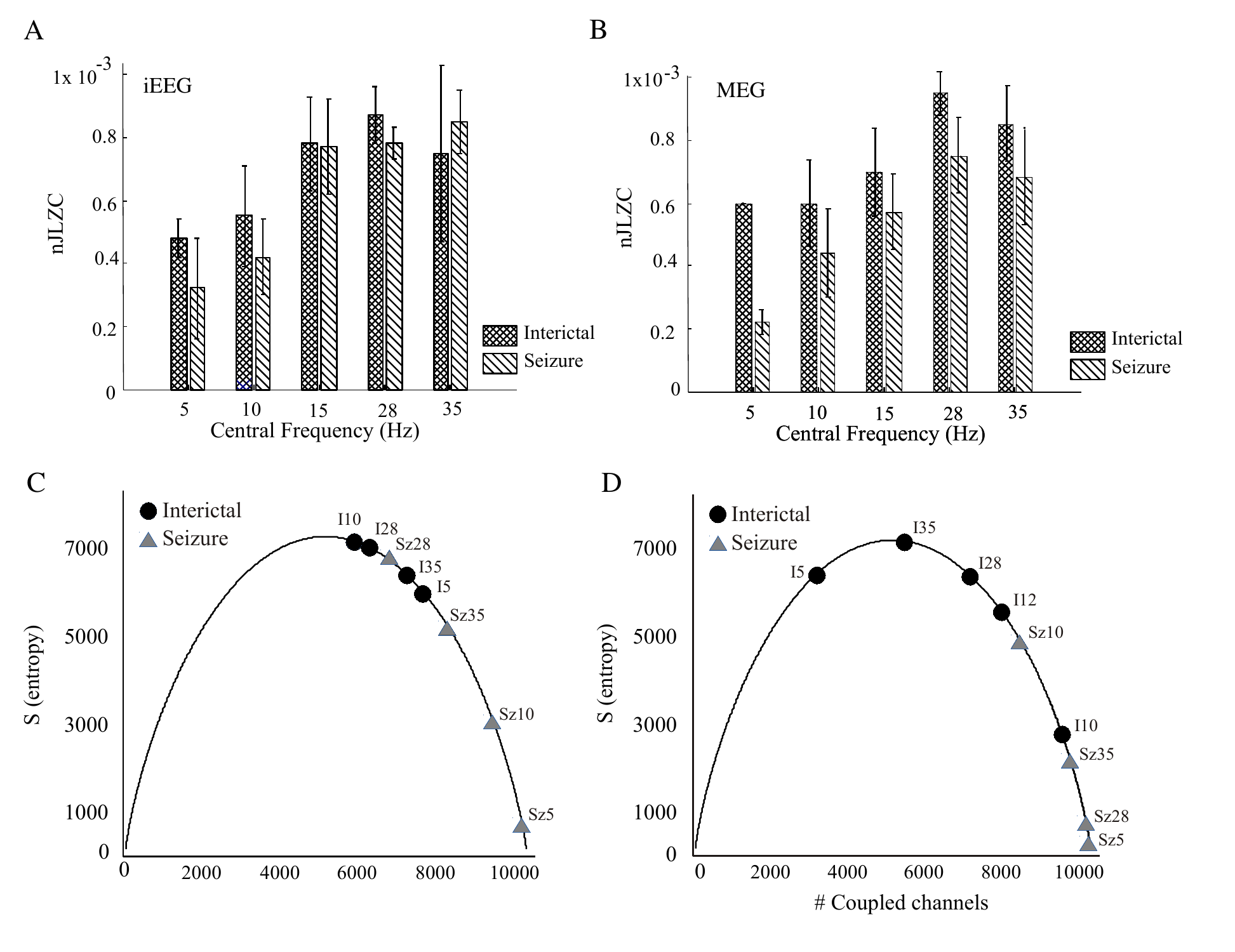} 
\caption{Normalized joint Lempel-Ziv complexity (nJLZC) and the entropy (S) for MEG and intracranial EEG (iEEG)  seizure recordings for different central frequencies (5,10,15,28,35 Hz). A) Mean nJLZC for 7 patients recorded with iEEG in interictal (cross  hatched bar) and ictal (seizure) (diagonal hatched bar) states. B) nJLZC  derived from  3 patients and 6 MEG recordings. Comparing A and B, it can be seen that lower complexity is associated with ictal states but the differences are more pronounced in case of MEG signals C,D) Entropy for one patient recorded with MEG and iEEG respectively, the dots and the triangles represent the interictal and the ictal (seizure) values for the different central frequencies. The Gaussian curve in these plots is described in Methods, and represents entropy values of all possible numbers of pairwise combinations of signals.  }
\label{fig:Figure2}
\end{figure}

Whereas the results using the macroscopic entropic measure and the microscopic nJLZC tend to be similar, there are important differences that provide insight into the brain dynamics in the distinct states. While the technical description of the entropy and the nJLZC is presented in Methods, here we remark in simple words what these measures mean. Basically, the entropic measure is that associated with the total number of configurations of connected brain networks (pairwise, as we use synchrony) in each experimental condition, thus the more possible configurations of connections the larger the entropy, which results in low entropy for either too many connections (as occurs in seizures, coma, and sometimes sleep) or too few as sometimes occurs in sleep (see reference 6 for a detailed discussion of this entropy). On the other hand, the nJLZC evaluates the fluctuations in the connectivity pattern of the entire combinations of networks (or, more precisely, signals) in each montage and in very short time windows. Hence, we have been able to set entropy and complexity measures in the context of connections among brain networks to evaluate the variability and the global information content of the system. In the cases of loss of consciousness due to pathological (non-physiological) states, both the entropy and the nJLZC have consistently lower values as compared with conscious states. Representative results are depicted in Figures \ref{fig:Figure1} and \ref{fig:Figure2} for the epileptic recordings. Most of the entropy results were already presented in \cite{erra2016statistical}, and here we just show some representative in order to compare the results with those obtained using the nJLZC. Figure \ref{fig:Figure1} shows the time course of the nJLZC as patients develop seizures. Importantly, the decrease in complexity is more manifest when the nJLZC is evaluated using signals over the whole brain, as opposed to only those from one hemisphere (Figure \ref{fig:Figure1}A). Importantly, Figure \ref{fig:Figure1}B illustrates the decrease in complexity (evaluated at 5 Hz) only after the patient lost consciousness,that occurred when the seizure generalised, at $350$ sec. Thus, this patient allowed us to correlate loss of consciousness with complexity, regardless of the presence of abnormal seizure activity in some brain areas. It was previous shown that, similarly, the entropy decreased only after the seizure generalised and patient lost consciousness (figure 1c in \cite{erra2016statistical}). Shown in the figure is the nLZC evaluated from the phase synchrony at $15$ Hz to demonstrate that the changes in complexity thus computed depends on the frequency at which synchrony is analysed. As discussed in the next section, this is not surprising due to the complex nature of synchronization even during robust ictal activity.

\begin{figure}
\includegraphics[scale=0.8]{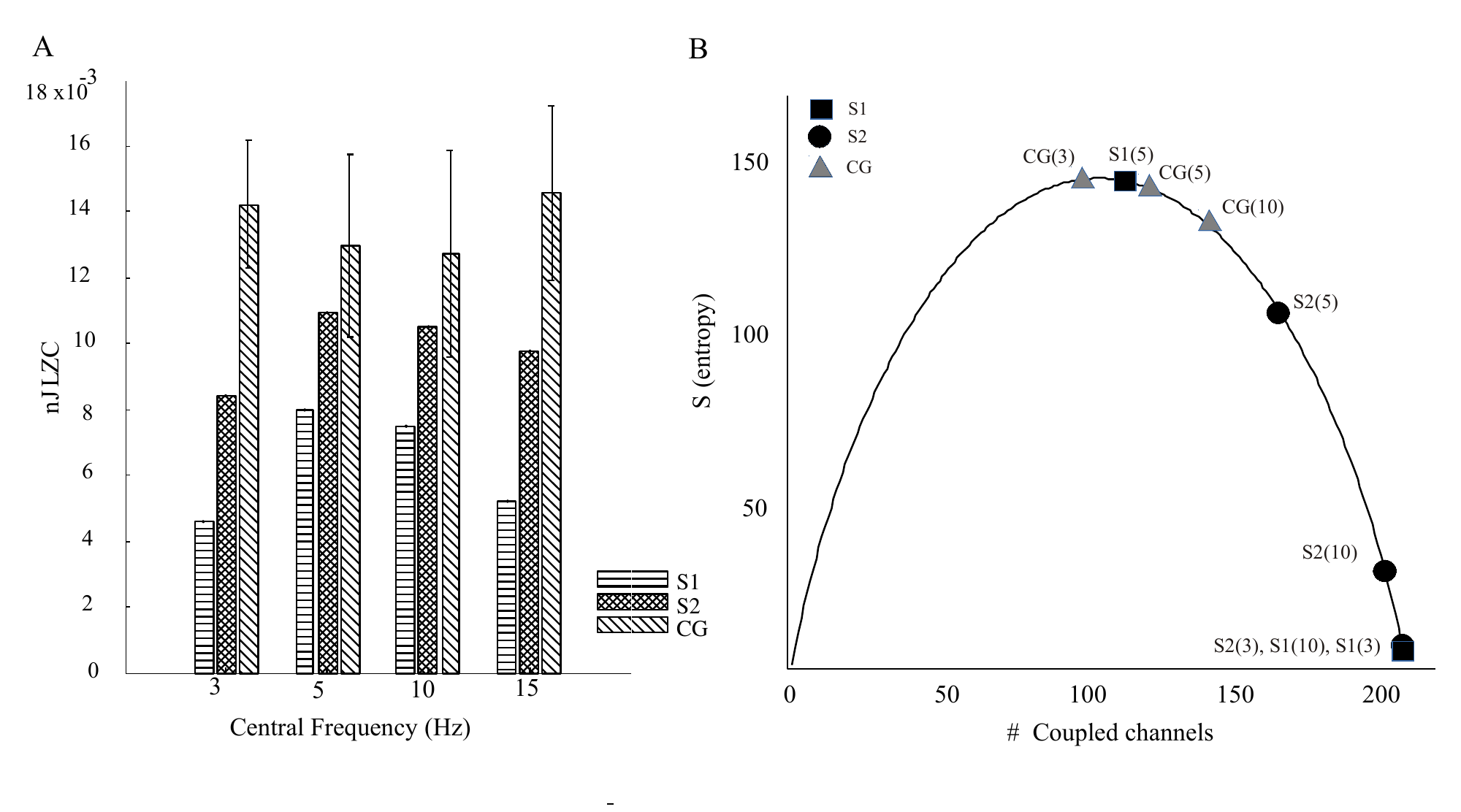} 
\caption{Normalized joint Lempel-Ziv complexity (nJLZC) (A) and entropy (S) (B) of scalp EEG recordings in two patients in coma (S1 and S2) and a control group (CG, 10 healthy subjects) for the central frequencies indicated. }
\label{fig:Figure3}
\end{figure}

Figure \ref{fig:Figure2} shows that generalized seizures (that normally imply loss of consciousness) are characterized by lower entropy and nJLZC. The bar graphs represent the average nJLZC for several patients recorded with iEEG or MEG. It is of interest to note that lower frequencies display largest differences, for both measures, between the conscious (interictal, or between seizures, brain activity) and the seizure state. This can be seen too in Table 1 and 2 in  Supplementary Information that presents all mean values of nJLZC for the different groups of epileptic patients, some recorded with MEG and others with iEEG; the same trend of lower complexity associated with ictal activity, most clearly at lower frequencies, can be appreciated, and as well that the differences between interictal and ictal states in the cases of iEEG are not as pronounced as those using MEG,  possibly due to the localized distribution of the intracerebral sensors, which renders this technique not as global as the MEG recordings that cover the whole cortex. These considerations are further elaborated and discussed below. In another pathological unconscious  state, Figure \ref{fig:Figure3} shows the entropy and nJLZC in coma, as compared to normal, conscious individuals, demonstrating lower nJLZC and entropy of the connectivity during coma.

\begin{figure}
\includegraphics[scale=0.68]{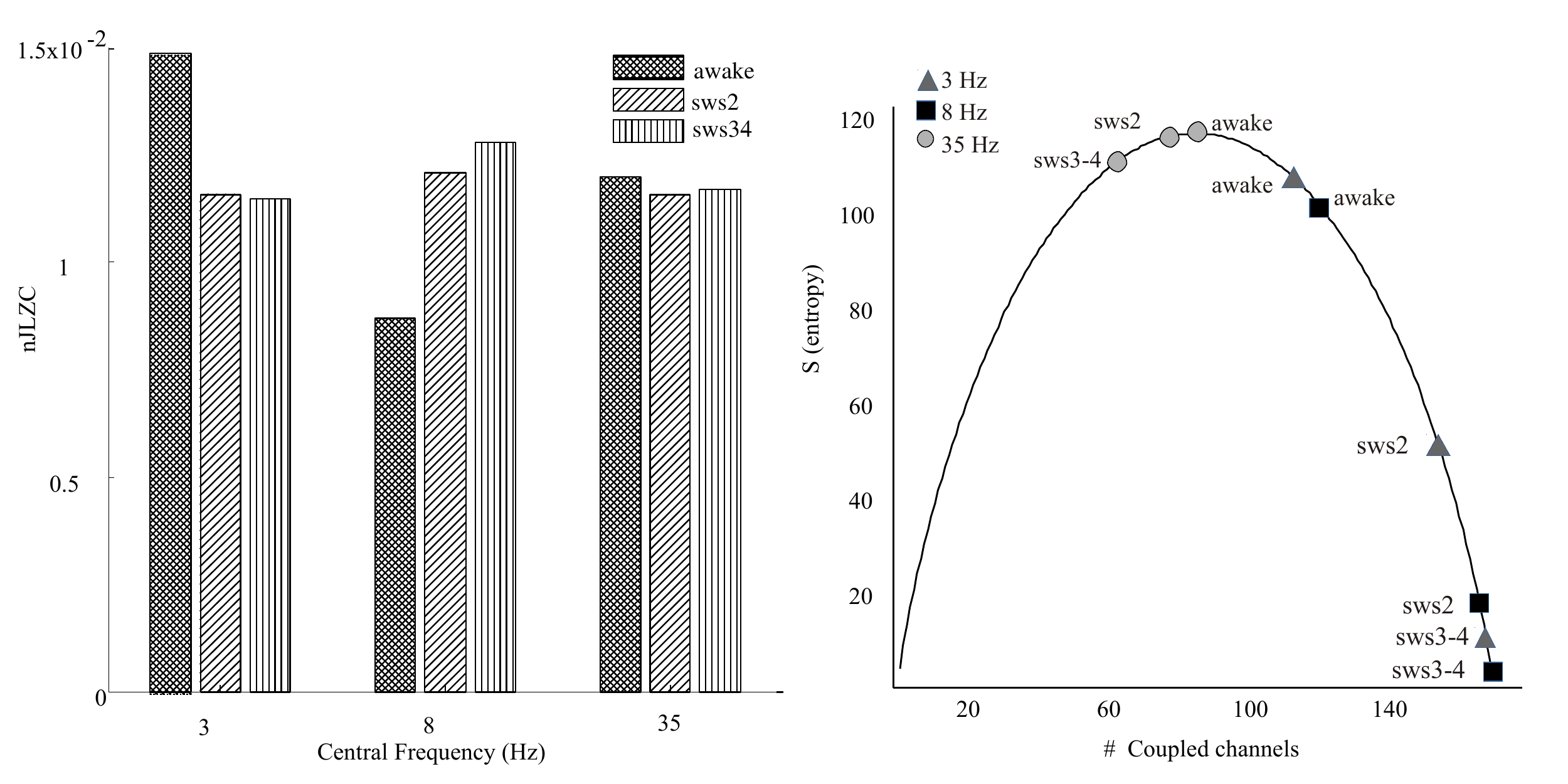} 
\caption{ Normalized joint Lempel-Ziv complexity (nJLZC) (A) and entropy (B) in one subject during different slow wave sleep (sws) stages (sws2 and sws3-4) and wakefulness, calculated at three    central frequencies of 3,8,35 Hz. Note that unlike nJLZC, the entropy tends to be higher in the awake state at all frequencies but more marked at lower ones. See text for details and discussion. }
\label{fig:Figure4}
\end{figure}

While the previously shown pathological unconscious states display same tendency towards lower entropy and nLZC, in the case of the physiological unsconscious states we analysed (sleep stages), there are important differences between the nJLZC and the entropy estimation. As illustrated for one representative case, Figure \ref{fig:Figure4} shows that, while the entropy was consistently lower during slow wave sleep (SWS) stages, as was found in \cite{erra2016statistical}, the nJLZC had lower values only at the lowest frequency studied (3 Hz). Tables 4 and 5 in Supplementary Information show all the values across all subjects in the different states and frequency bands, where it is apparent that it is at low frequencies (4-12 Hz) when the nJLZC is consistently lower than in fully alert states, and the differences are greater between the deepest SWS stage (SWS3-4) and the awake state. It is also of note that, in those subjects where rapid-eyes movement (REM) states could be recorded, the nJLZC is similar or even higher than the awake state, a result that parallels that found in the entropy values reported in \cite{erra2016statistical}.   

\section{Discussion}

In summary, our observations suggest that conscious awareness is associated with maximal entropy that represents a global, macroscopic perspective of the total number of configurations of connected brain networks; on the other hand, the more local, microscopic view of the fluctuations in connectivity patterns given by the nJLZC does not consistently show a decreased complexity,and this occurs especially during sleep stages where there could be unconscious cognitive processes. Hence, conscious awareness seems to be best described as a global brain state. As such, our results support not only the main theme underlying several major theories of consciousness and cognition, namely, the requisite for a substantial number of microstates  -- or configurations of neuronal connections, that is, variability in the establishment of functional connected networks -- in order to increase the integrated information (as postulated by the information integrated theory \cite{tononi2004information}), to broadly broadcast activity to many different cell networks (the global workspace theory proposes that widespread distribution of information leads to conscious awareness \cite{baars1993cognitive}), and to avoid becoming trapped in one stable activity pattern (metastability \cite{kelso1997dynamic}), but also endorse numerous computational and theoretical studies indicating that the variability in the patterns of organised activity arising from the maximization of fluctuations in synchrony \cite{vuksanovic2015dynamic} is fundamental for a healthy brain \cite{garrett2013moment}. It is fair to remark that, while we have used the term ‘connectivity’, in reality our synchrony analysis reveals only correlation between phases of oscillation, as already discussed in Methods; nevertheless, some basic assumptions have to be made and it is reasonable to infer that these correlations of activity may underlie some degree, or probability, of connectivity. In this regard, our results are consistent with the computational observation that transitions between conscious states are achieved by just varying the probability of connections in neural nets \cite{zhou2015percolation}.

These observations indicate that the measures of entropy and nJLZC as have been used here are associated with the handling of information, or cognitive processing, that may or may not be related to conscious awareness. Thus, we find that while our macroscopic measure of entropy  consistently distinguishes conscious and unconscious states, the analysis of the microscopic fluctuations at short time scales of the configurations using nJLZC reveals high complexity during apparent unconscious states like SWS and REM. But this is not really surprising, as substantial brain activity has been demonstrated not only during REM but also during non-REM sleep using a variety of methods, including fMRI and electrophysiological recordings \cite{achermann1998coherence,stickgold2001watching,rees2002neural,wehrle2007functional}, studies that indicated that the oscillatory activity during non-REM phases were more complex than previously thought and that there is a re-organization of brain networks into localised modules during SWS \cite{spoormaker2012frontoparietal}. Additionally, theoretical studies have assessed first-order phase transitions during the sleep cycle that bring forth fluctuations in activity \cite{steyn2005sleep}. Brain activity during sleep is theorised to be necessary for memory consolidation \cite{sejnowski2000we,demanuele2016coordination}.

As to the epileptiform activity here studied, a decrease in entropy was previously reported specifically in generalised seizures when there is a loss of conscious awareness \cite{erra2016statistical}, and our nJLZC studies here show the same phenomenon. Moreover, we had the advantage of one of our patients experiencing generalization of the seizure at a later stage and thus remaining conscious during the beginning of the ictus (Figure 1B). The reduced complexity during seizures is less clear as the recording sites become more localised, as shown in our analysis of iEEG signals, which substantiates the global nature of consciousness. It should be noted too that loss of consciousness during seizures is variable and normally coincides with the degree of generalization \cite{blumenfeld2003seizures}. Once again, then, as opposed to the global entropy computation, the nJLZC captures not only the microstate fluctuations during sleep but also the intricacy of the complexity of synchrony patterns in seizures, in keeping with other works that have shown the multi-scale processes of synchronization in epilepsy \cite{chavez2005intrinsic,amor2009cortical,filatov2011dynamics}.

Considering the reports that have observed a repertoire of ongoing brain states and common motifs of connections in resting conditions \cite{betzel2012synchronization,steinke2011brain,chu2012emergence}, it is tempting to conjecture that nJLZC captures this ongoing recurrent activity that is a feature of the normal, healthy brain, but that is lost during pathological conditions when functional coordination is altered; hence, it is in these pathological cases (seizures, coma) when nJLZC drops consistently low, but not in sleep. Thus, our measures may be more related to reflect optimality of nervous system dynamics, rather than consciousness as such. Because these fluctuations of spontaneous activity have been proposed to constitute a fundamental principle of brain organization \cite{pinsk2007neuroscience}, our observations help frame in a more formal sense the inquiry into how consciousness arises from the organization of nervous system tissue. We also advance, in view of our results that the entropy and complexity of slow oscillations differentiate best conscious and unconscious states, that the low frequency oscillations represent a most fundamental property of the underlying (self)organization of cell assemblies, and indeed it was recently  proposed that “slow oscillations provide information about the underlying healthy or pathological network” \cite{sanchez2017shaping}. The communication among neurons via nested rhythms may require synchronization at low frequencies that serves as temporal reference for information transfer at other, higher frequencies observed during wakefulness \cite{baptista2008transmission,bonnefond2017communication}. Our studies thus may contribute to the current debate on the possible, putative roles of gamma rhythms in cognition \cite{merker2016cortical}. To integrate these high-level perspective results with basic cellular properties will require more investigations but it is noteworthy that neurons possess a characteristic low-pass filter behaviour (thus favouring slow oscillations) that is lost after disturbances common to epilepsy or traumatic brain injury \cite{frantseva1998changes}.

The macroscopic entropy of the number of possible configurations we calculate here and in \cite{erra2016statistical} measures the information content of the functional network \cite{rieke1999spikes}. This framework of relating information to the organization of cognition may be very useful as other authors have already emphasised -- a classical text on processes of organization described by change of information content with time is Atlan \cite{atlan1974formal}. Previous work proposed that a general organising principle of natural phenomena is the tendency toward maximal, more probable, distribution of energy/matter \cite{velazquez2009finding}, a proposal that is encapsulated by the notion of maximization of information transfer \cite{smith2008thermodynamics}, thus we venture that the brain organization optimal for conscious awareness will be a manifestation of the tendency towards a maximal information exchange. Hence it is not surprising that when the organism is fully involved in processing sensory information this information is largest. On the other hand, even in moments of unconsciousness there can be substantial processing, which is revealed upon a closer scrutiny at the microscale level, as that provided by the JLZC. The results provide evidence for the notion that ongoing transformations of information in the brain are reflected in the variability and fluctuations in the functional connection among brain cell ensembles (large entropy of the number of possible configurations and concomitant large complexity), which manifest in aspects of consciousness.  The crucial aspect for a healthy brain dynamics then is not to reach maximum number of units interacting, but rather the largest possible number of configurations (allowed by the constraints).  In this regard, the maximum entropy approach has been advocated as a simplifying framework for the study of networks \cite{stephens2011searching}. Thus, in conclusion, our studies shed light on two levels: at a conceptual level they emphasise the global nature of conscious awareness, whereas at a more practical level, perhaps of use in the clinic, they reveal information processing that can occur unconsciously. We thus hope that our studies will help find general principles of the structure of cognition that underlie aspects of conscious awareness, following the advice of S. Dehaene: ``much [$ \ldots $] has focused on the details of a few specific phenomena, rather than on the general architecture of cognition'' \cite{dehaene2007few}.


\subsubsection*{Acknowledgment}
Our work is supported by a Discovery grant from the Natural Sciences and Engineering Research Council of Canada (NSERC).



\bibliography{JLZC_Bibliography}
\bibliographystyle{unsrt}

\end{document}